\documentclass{iaus}
\usepackage{graphicx}

\title[X-ray spectral diagnostics] %% give here short title %%
{X-ray spectral diagnostics of activity in massive stars}

\author[David H.\ Cohen, Emma E.\ Wollman, \& Maurice A. Leutenegger]   %% give here short author list %%
{David H.\ Cohen$^1$, Emma E.\ Wollman$^2$, \& Maurice A. Leutenegger$^3$}

\affiliation{$^1$Department of Physics and Astronomy, Swarthmore College,\\ 500 College Ave., Swarthmore, Pennsylvania, 19081, USA \\ email: {\tt cohen@astro.swarthmore.edu} \\[\affilskip]
$^2$Department of Physics, California Institute of Technology,\\ Pasadena, California, 91125, USA \\email: {\tt ewollman@caltech.edu} \\[\affilskip]
$^3$NASA/Goddard Spaceflight Center, Code 662, Greenbelt, Maryland, 20771, USA \\ email: {\tt maurice.a.leutenegger@nasa.gov}
}

\pubyear{2010}
\volume{272}  %% insert here IAU Symposium No.
\pagerange{--}
\jname{Active OB Stars}
\editors{C. Neiner, G. Wade, G. Meynet, \& G. Peters, eds.}
\begin{document}

\maketitle

\begin{abstract}
  X-rays give direct evidence of instabilities, time-variable
  structure, and shock heating in the winds of O stars. The observed
  broad X-ray emission lines provide information about the kinematics
  of shock-heated wind plasma, enabling us to test wind-shock models.
  And their shapes provide information about wind absorption, and thus
  about the wind mass-loss rates.  Mass-loss rates determined from
  X-ray line profiles are not sensitive to density-squared clumping
  effects, and indicate mass-loss rate reductions of factors of 3 to 6
  over traditional diagnostics that suffer from density-squared
  effects.  Broad-band X-ray spectral energy distributions also
  provide mass-loss rate information via soft X-ray absorption
  signatures.  In some cases, the degree of wind absorption is so
  high that the hardening of the X-ray SED can be quite significant.
  We discuss these results as applied to the early O stars $\zeta$ Pup
  (O4 If), 9 Sgr (O4 V((f))), and HD 93129A (O2 If*).  \keywords{line:
    formation, shock waves, stars: winds, x-rays: stars}

\end{abstract}

\firstsection % if your document starts with a section,
\section{Introduction}

Soft X-ray emission is ubiquitous in O stars, and it is generally
accepted that it arises in numerous shock-heated regions embedded in
these stars' powerful and dense radiation-driven winds. The broad
emission lines seen in high-resolution X-ray spectra of O stars
confirm this scenario qualitatively.  In this paper, we present
quantitative analysis of resolved X-ray emission lines observed in
three very early O stars, from which we are able to place constraints
on the kinematics and spatial distribution of the shock-heated plasma
and thereby test predictions of numerical simulations of wind shocks.
We also show how the degree of attenuation by the bulk wind in which
the shocked plasma is embedded can be measured both from resolved
emission lines and from lower resolution broadband X-ray spectra in
order to estimate the mass-loss rates of O star winds.

We restrict our discussion to ``normal'' massive stars, where binarity
and the associated colliding wind shock (CWS) X-ray emission and
magnetically channeled wind shock (MCWS) X-ray emission is absent or
negligible. The dominant paradigm for X-ray production in normal O and
early-B stars is the embedded wind shock (EWS) scenario, and the
specific mechanism for EWSs is usually assumed to involve the
line-driving instability (LDI), either in a self-excited mode
(\cite{ocr1988}) or in a mode where the instability is seeded by
perturbations at the base of the wind (\cite{Feldmeier1997}). 

\begin{figure}[b]
\vspace*{-0.5 cm}
\begin{center}
 \includegraphics[angle=90, scale=0.15]{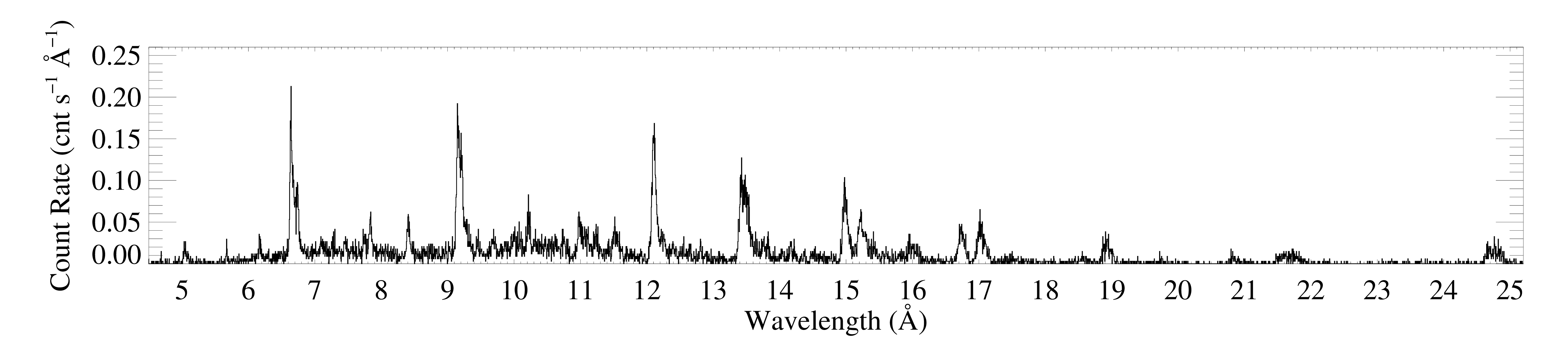} 
\vspace*{-0.25 cm}
 \includegraphics[angle=90, scale=0.15]{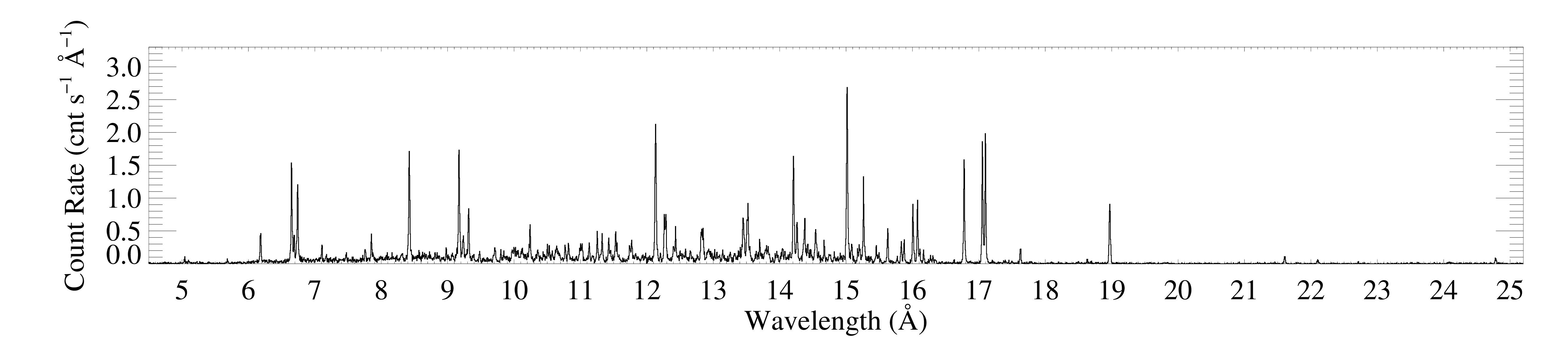} 
%\vspace*{-0.5 cm}
 \caption{{\it Chandra} medium energy grating (MEG) spectra of the O4
   If star, $\zeta$ Pup (top) and for comparison, of the G star,
   Capella (bottom). The spectrum of the O star is harder (strongest
   lines at shortest wavelengths) but by comparing H-like and He-like
   line strengths (e.g.\ of Mg at 8.42 \AA\/ and 9.2 \AA,
   respectively) it is evident that the higher temperature plasma is
   found on the G star.  Finally, note that the emission lines are
   unresolved in the Capella spectrum and are significantly broadened
   in the $\zeta$ Pup spectrum. }
   \label{fig:broadband}
\end{center}
\end{figure}

The morphology of high-resolution X-ray spectra of normal massive
stars reveals some important qualitative properties of O star X-rays.
In Fig.\ \ref{fig:broadband} we compare the O supergiant $\zeta$ Pup
to the coronal G star, Capella, to highlight some of these properties.
The O star's spectrum is harder, overall, than the G star's, however
this is due not to higher plasma temperatures, but rather to the
effects of wind absorption, consistent with the X-rays arising in the
dense stellar wind of the O star. This broadband view of the X-ray
spectra also shows quite obviously that the emission lines in the O
star are much broader than the (unresolved) lines in the G star, as
the EWS scenario predicts. In the next section, we show how
quantitative information can be derived from the Doppler broadened
X-ray emission lines.

\section{The X-ray line profile model applied to $\zeta$ Pup}

To extract information from individual resolved line profiles, we fit
a simple wind-shock model informed by the LDI simulations, in which
numerous shock-heated regions are distributed throughout the wind
above some shock onset radius, $R_{\rm o}$, with local emission
measure assumed to scale with the local ambient wind density squared
(\cite{oc2001}). The kinematic profile of the X-ray plasma is assumed
to trace the same beta-velocity law that describes the bulk wind.
This assumption is based on the results of numerical simulations of
EWSs that show accelerated pre-shock wind streams being decelerated
back down the local ambient wind velocity (\cite{ro2002}). The
attenuation due to continuum opacity in the bulk wind in which the
shock-heated plasma is embedded is described by the characteristic
optical depth parameter, $\tau_{\ast} \equiv
\kappa{\dot{M}}/4{\pi}R_{\rm \ast}v_{\infty}$, where $\kappa$ is the
(wavelength dependent) opacity of the bulk wind, $\dot{M}$ is the wind
mass-loss rate, $R_{\rm \ast}$ is the stellar radius, and $v_{\infty}$
is the wind terminal velocity. The absorption of X-rays imparts a
characteristic blue-shifted and asymmetric shape to the emission line
profiles due to the preferential attenuation of red-shifted line
photons emitted in the far hemisphere of the wind, while leaving
blue-shifted line photons emitted from the near hemisphere much less
attenuated.

\begin{figure}[b]
% \vspace*{-2.0 cm}
\begin{center}
 \includegraphics[angle=90, scale=0.19]{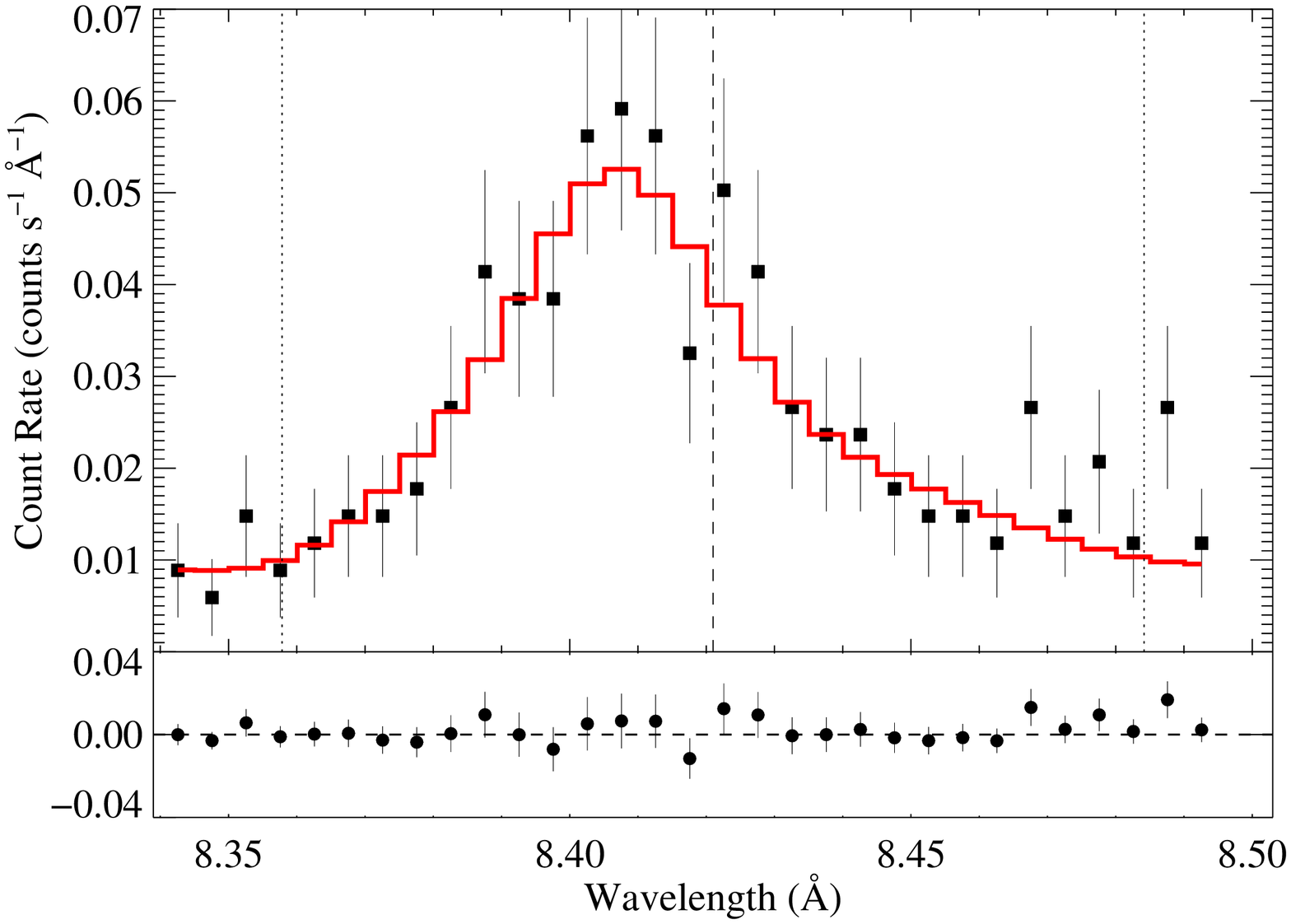} 
 \includegraphics[angle=90, scale=0.19]{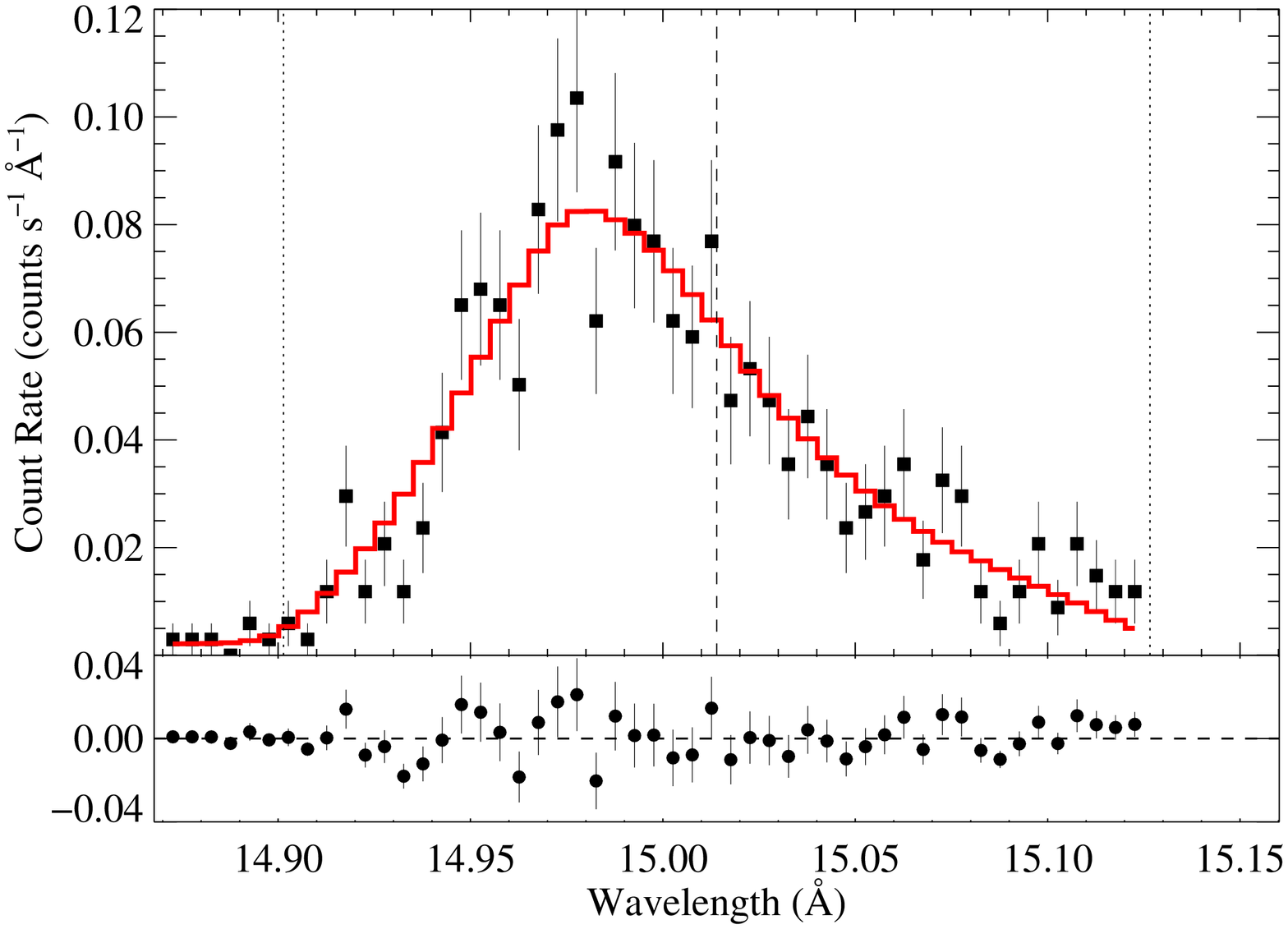} 
 \includegraphics[angle=90, scale=0.19]{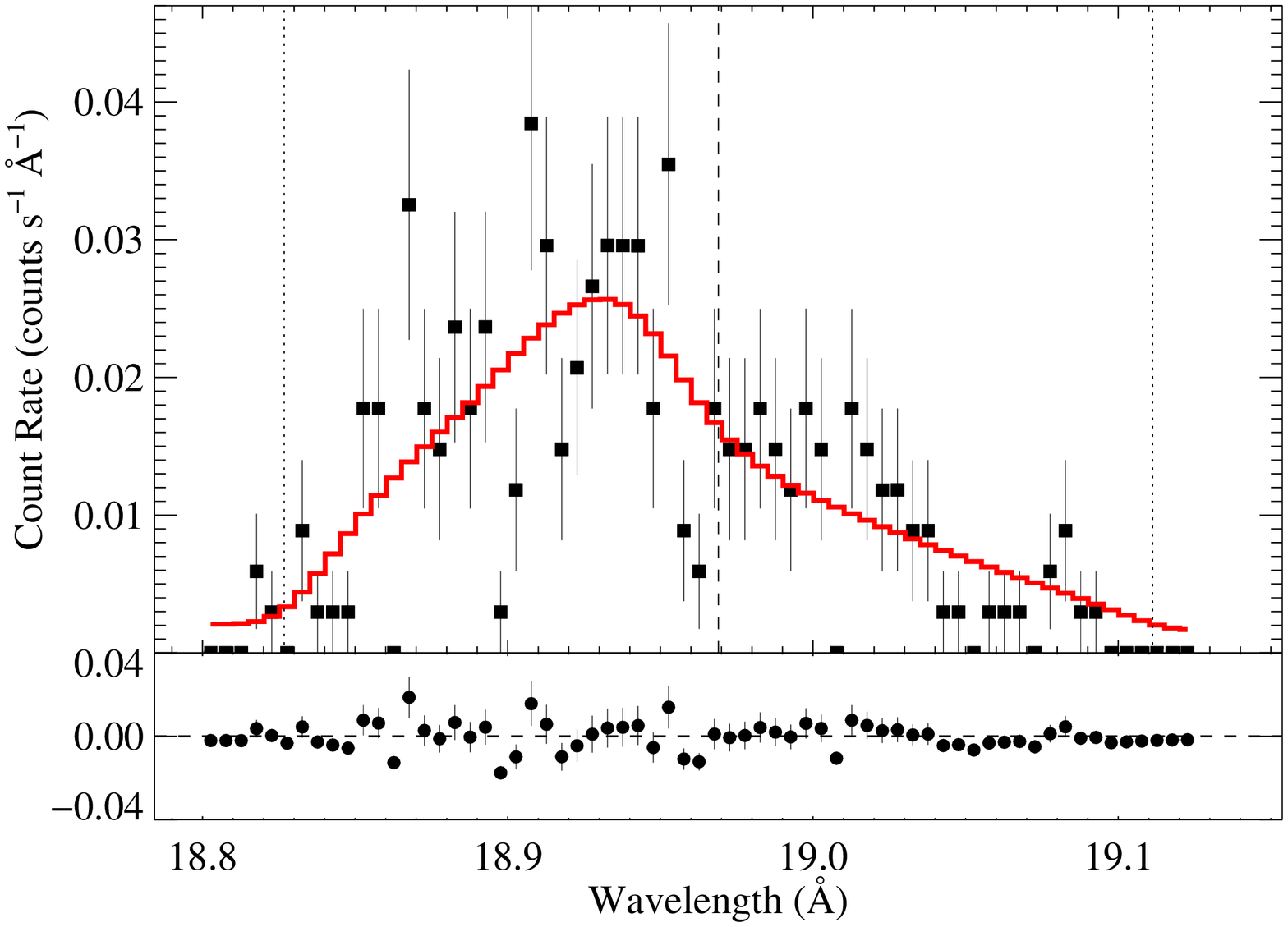} 
% \vspace*{-1.0 cm}
 \caption{Fits to three lines in the {\it Chandra} spectrum of $\zeta$
   Pup.  From left to right: the Ly$\alpha$ line of Mg\, {\sc xii} at
   8.42 \AA, the Fe\, {\sc xvii} line at 15.01 \AA, and the Ly$\alpha$
   line of O\, {\sc viii} at 18.97 \AA. The vertical dashed lines in
   each panel represent the laboratory rest wavelengths of each
   transition, while the flanking dotted lines represent the Doppler
   shifts associated with the wind terminal velocity.  The
   characteristic broad, blue shifted, and asymmetric profile shapes
   are evident, as is an increase in the shift and asymmetry with
   wavelength, as is expected from the form of the continuum opacity
   of the bulk wind, which generally increases with wavelength.  The
   characteristic optical depths of these three lines are roughly
   $\tau_{\ast} = 1, 2$, and 3, respectively. }
   \label{fig:zpup_fits}
\end{center}
\end{figure}

\begin{figure}[b]
%\vspace*{-1.4cm}
\begin{center}
 \includegraphics[angle=90, scale=0.19]{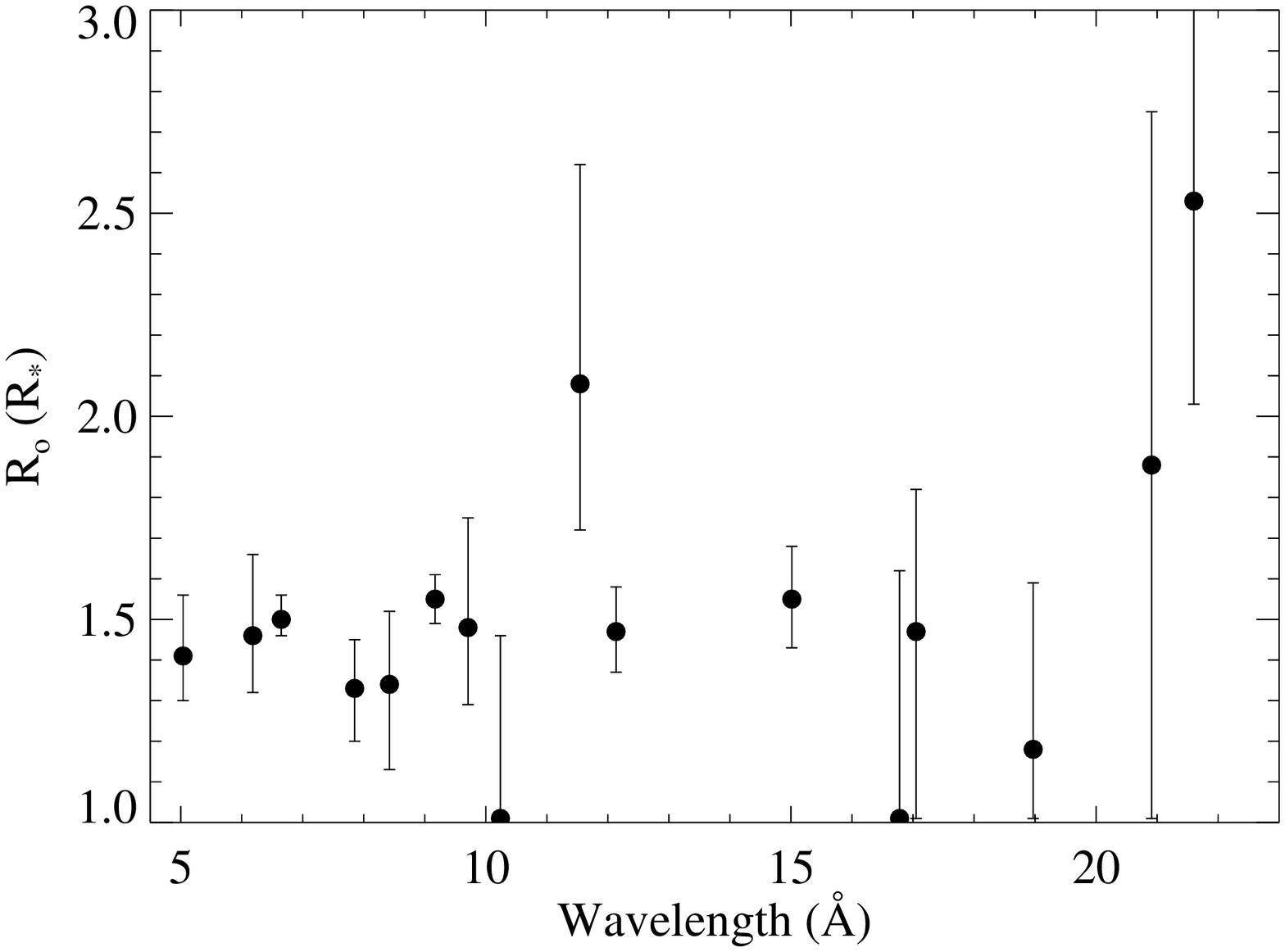} 
 \includegraphics[angle=90, scale=0.19]{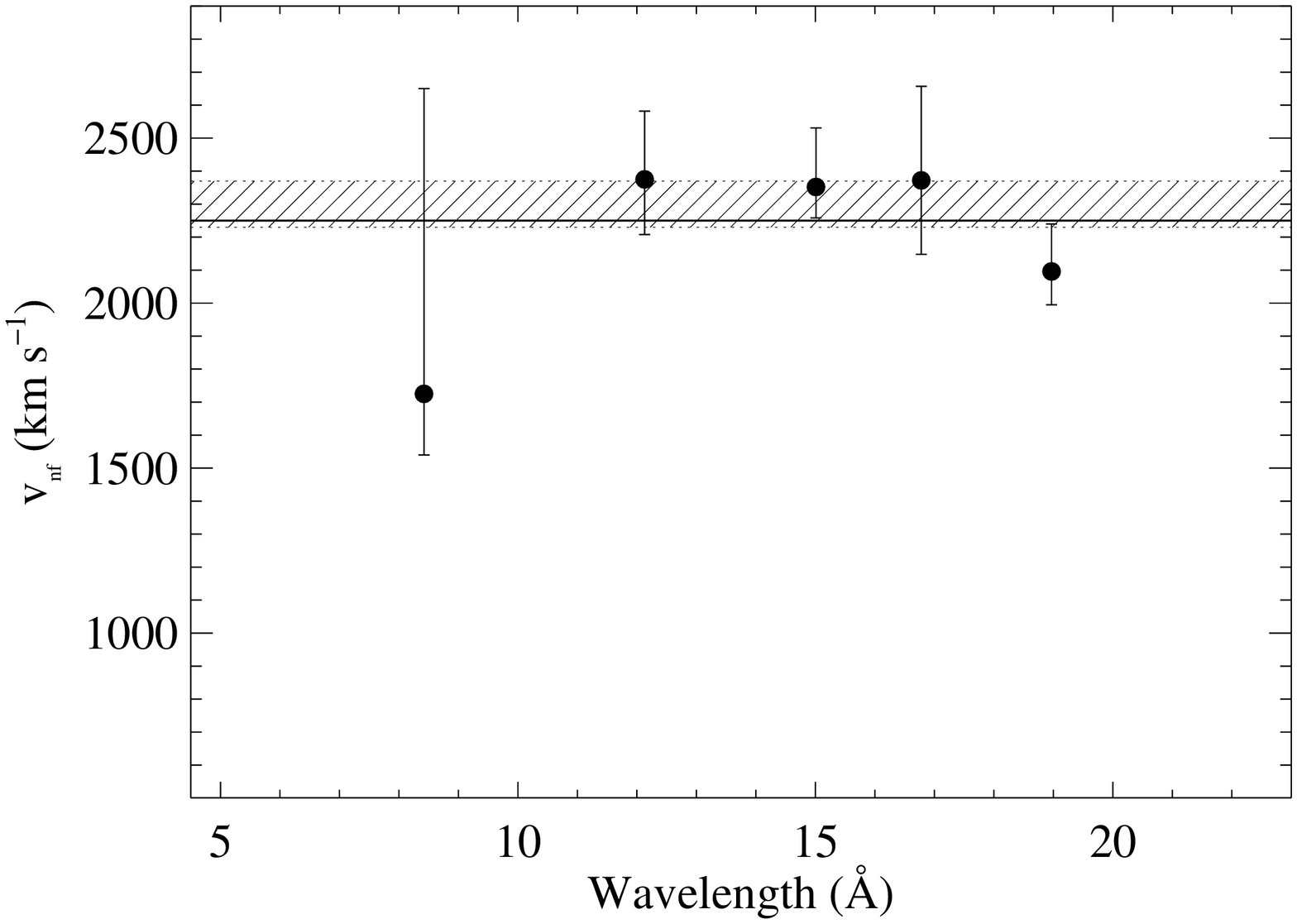} 
 \includegraphics[angle=90, scale=0.19]{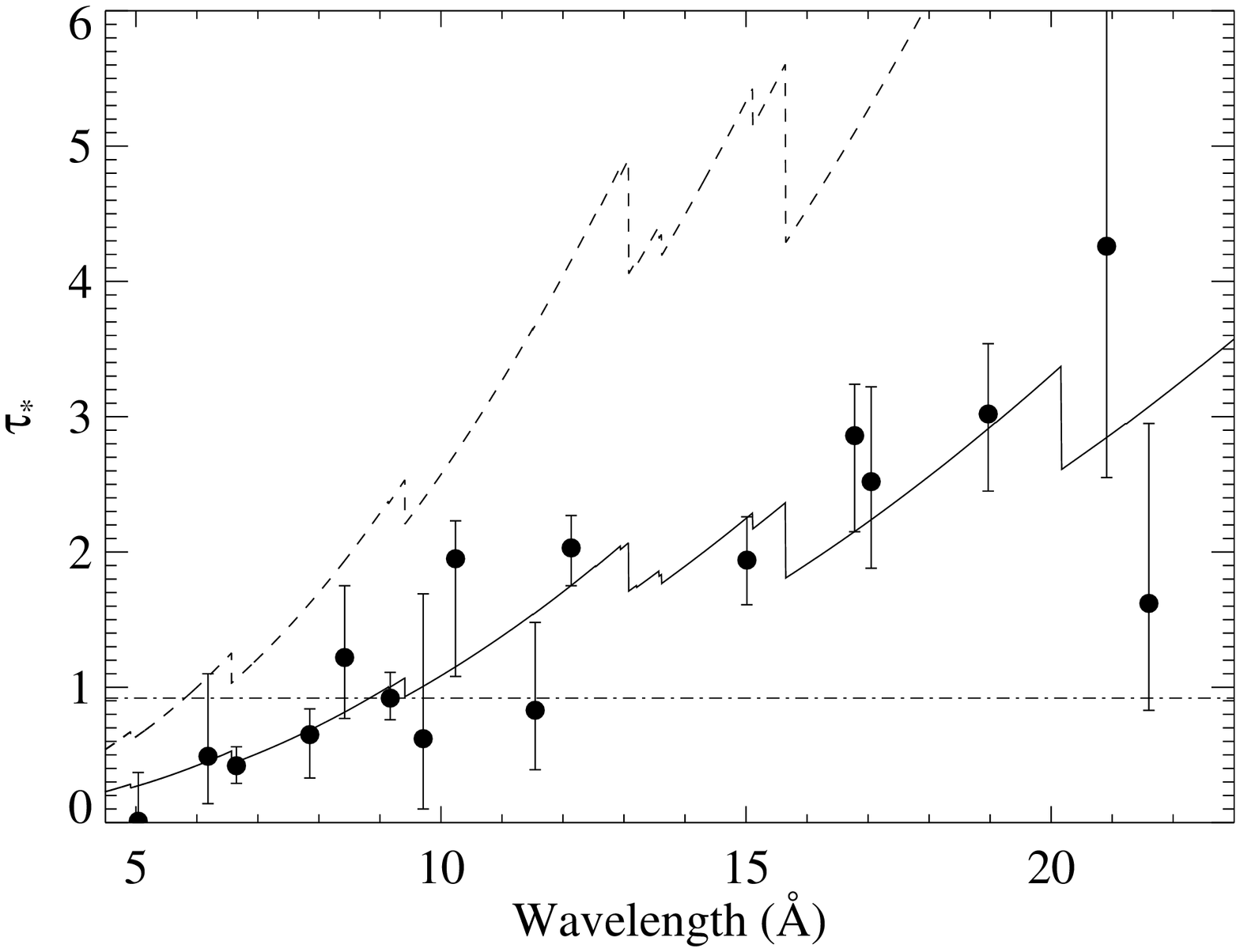} 
 \caption{Results from fitting the wind-profile model to the emission
   lines in the {\it Chandra} spectrum of $\zeta$ Pup.  From left to
   right: the shock onset radii, $R_{\rm o}$, which are consistent
   with a universal value of 1.5 R$_{\ast}$; the wind terminal
   velocities for the five strongest, unblended lines in the spectrum,
   which are consistent with the value for the bulk wind, of
   $v_{\infty} = 2250$ km s$^{-1}$, derived from UV spectra (this
   value is represented by the horizontal line, while the cross
   hatched region is the 68\% confidence limit on the mean value of
   the five fitted terminal velocities shown as points with error
   bars); and the $\tau_{\ast}$ values from each of the 16 fitted line
   profiles.  This last panel shows that a constant value of the
   characteristic optical depth provides a poor fit, as does a model
   that incorporates the continuum opacity of the bulk wind but
   assumes a high value for the mass-loss rate ($8.3 \times 10^{-6}$
   ${\rm M_{\odot}~{\rm yr^{-1}}}$ ), while a model with the mass-loss
   rate as a free parameter provides a good fit, with a mass-loss rate
   of $3.5 \times 10^{-6}$ ${\rm M_{\odot}~{\rm yr^{-1}}}$.}
   \label{fig:zpup_results}
\end{center}
\end{figure}

For each emission line in the {\it Chandra} spectrum of an O star, we
can fit this empirical profile model and derive best-fit values of
$R_{\rm o}$ and $\tau_{\rm \ast}$ by minimizing the C statistic, and
place confidence limits on them via the $\Delta{\chi}^2$ formalism
applied to the C statistic. For $\zeta$ Pup (summarizing the results
published in \cite[Cohen et al.\ (2010)]{Cohen2010}), we find -- for
16 lines and line complexes in the {\it Chandra} grating spectrum
(three representative lines and their best-fit profile models are
shown in Fig.\ \ref{fig:zpup_fits}) -- a universal value for the
shock-onset radius of $R_{\rm o} \approx 1.5$ ${\rm R_{\ast}}$, which
is consistent with numerical simulations of the LDI
(\cite{Feldmeier1997,ro2002}). We also find a range of characteristic
optical depths, $\tau_{\rm \ast}$, for the 16 emission lines,
consistent with the expected wavelength trend in the atomic opacity.
By calculating a detailed opacity model, and assuming standard values
for the stellar radius and wind terminal velocity, we fit the ensemble
of characteristic optical depths to find a best-fit mass-loss rate,
via $\dot{M} = 4{\pi}R_{\rm
  \ast}v_{\infty}\tau_{\ast}({\lambda})/\kappa({\lambda})$. The values
of the onset radius, $R_{\rm o}$, and of the characteristic optical
depths, $\tau_{\ast}$, are shown in Fig.\ \ref{fig:zpup_results}.  The
panel with the $\tau_{\ast}$ values also shows the best-fit model of
the wavelength-dependent optical depths, from which we derive a
mass-loss rate of $3.5 \pm 0.3 \times 10^{-6}$ ${\rm M_{\odot}~{\rm
    yr^{-1}}}$.

We emphasize that the mass-loss rate determination from the X-ray
profiles represents a factor of roughly three reduction from the
traditional H$\alpha$-derived mass-loss rate that ignores clumping
(\cite{Markova2004}).  And that this modest reduction in the mass-loss
rate is consistent with newer determinations using H$\alpha$ and radio
and IR free-free excesses that {\it do} account for clumping
(\cite{Puls2006}). Also, we note that for a high signal-to-noise {\it
  Chandra} spectrum with many emission lines, like that of $\zeta$
Pup, the statistical error on the derived mass-loss rate is small
(about 10\%), but that the actual uncertainty is dominated by
uncertainty in the wind opacity model, which in turn is dominated by
uncertainty in the elemental abundances.  Individual elemental
abundances do not have a large effect, but the overall metallicity
does.  The model we use in this paper (and which was used in
\cite[Cohen et al.\ (2010)]{Cohen2010}) uses subsolar metallicity and
C, N, and O abundances altered by CNO processing.  If future abundance
determinations are made which supersede the current ones, the
mass-loss rate should be rescaled in inverse proportion to the
metallicity adjustment (more metals cause higher opacity which would
then require lower wind column densities and so lower mass-loss
rates).

\section{Other O stars}

We can apply the same type of line profile analysis to other O stars
observed with the {\it Chandra} grating spectrometer. Here we present
preliminary analysis of the early O main sequence star, 9 Sgr, at the
center of the Lagoon Nebula, and the very early O supergiant, HD
93129A, in Tr 14 in Carina.  Both stars have binary companions, but in
neither case are the emission lines in the grating spectrum significantly
contaminated by the harder emission associated with CWB X-rays. 

\begin{figure}[b]
\begin{center}
 \includegraphics[angle=90, scale=0.19]{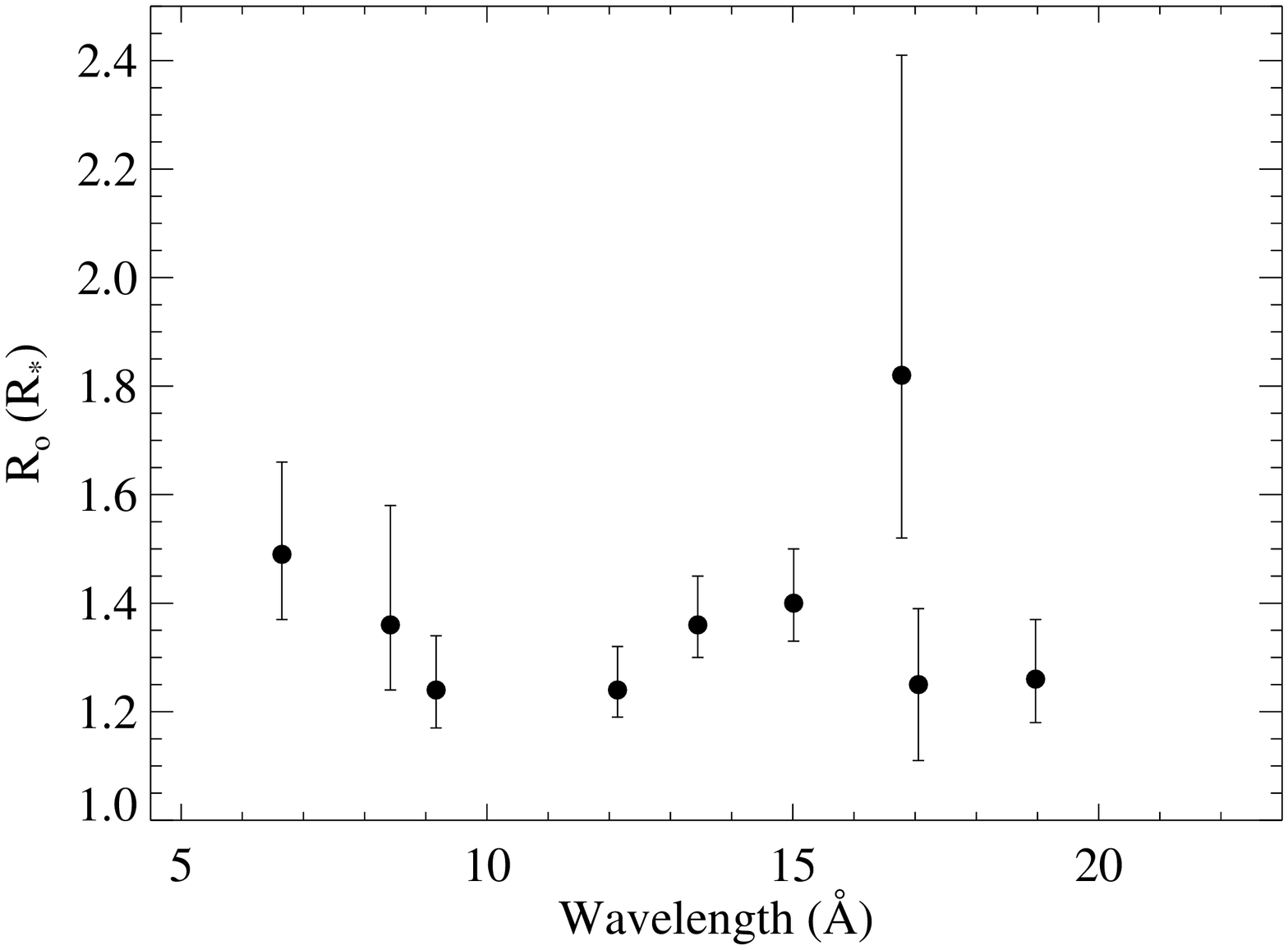} 
 \includegraphics[angle=90, scale=0.19]{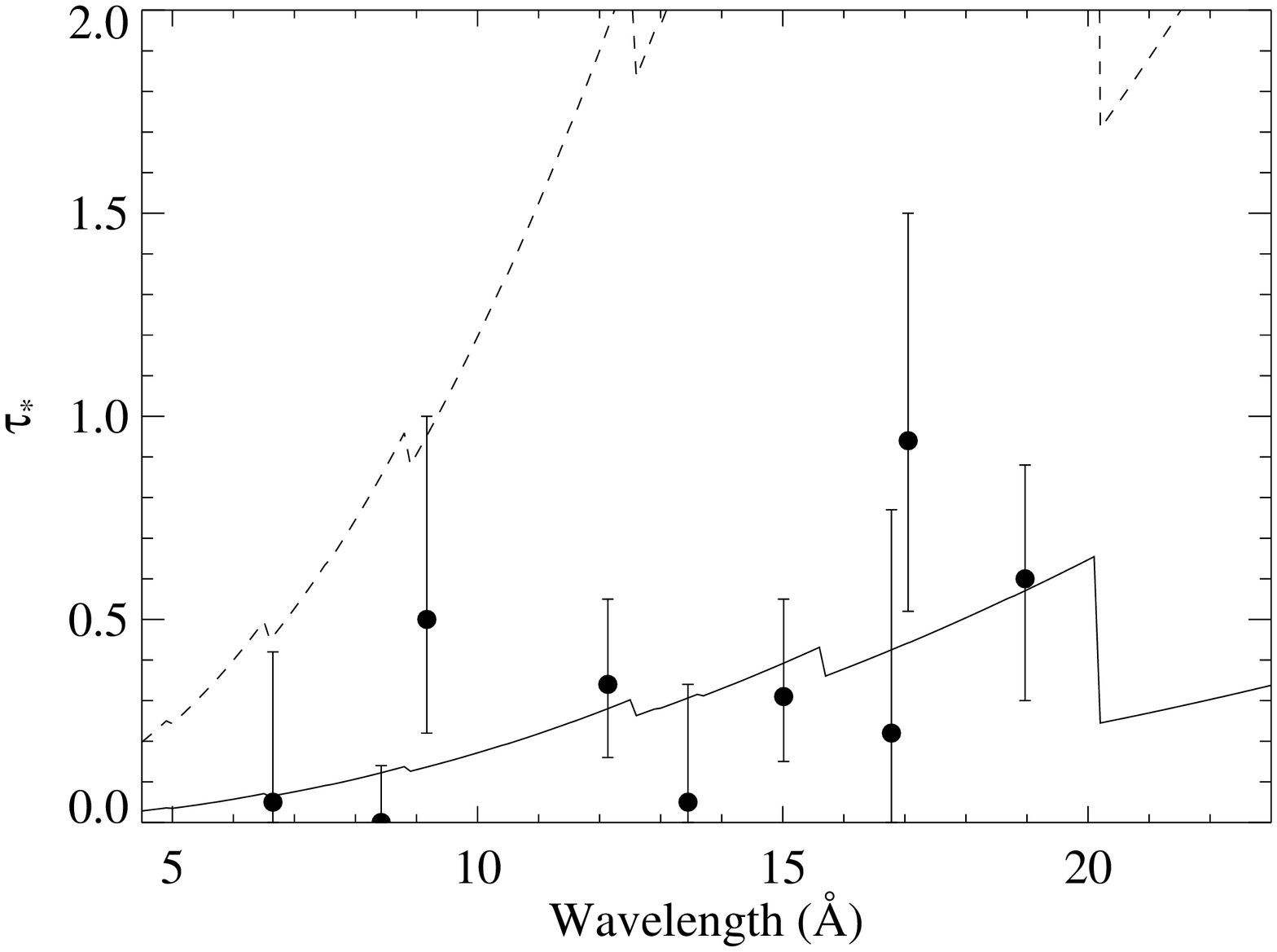} 
 \includegraphics[angle=90, scale=0.19]{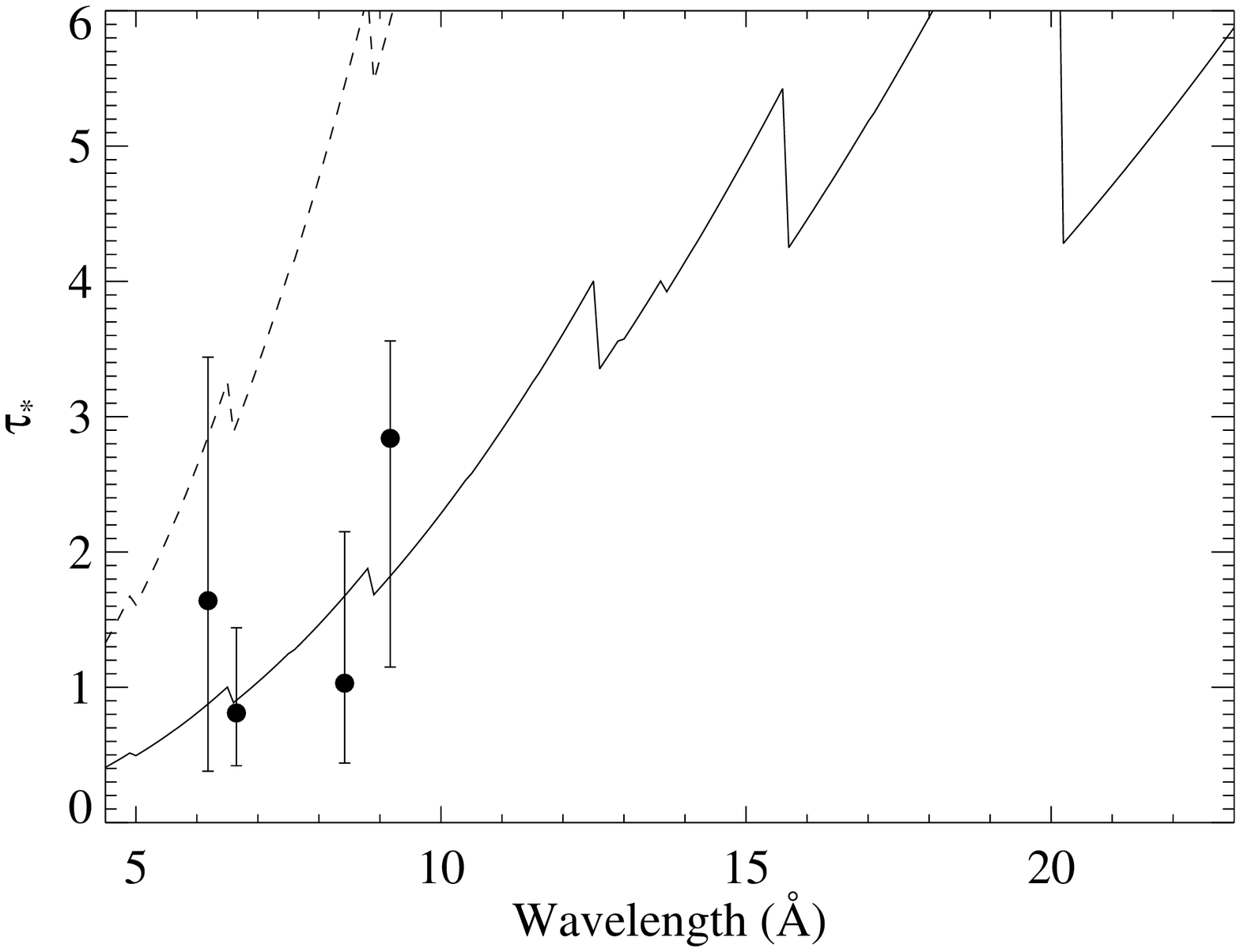} 
 \caption{Results from fitting the wind-profile model to the emission
   lines in the {\it Chandra} spectrum of 9 Sgr and HD 93129A. The
   shock onset radii for 9 Sgr (left) are consistent with a value of
   $R_{\rm o} = 1.4$ R$_{\ast}$, while the characteristic optical
   depths (center) are well fit by a model that has a mass-loss rate
   of $\dot{M} = 3.4 \times 10^{-7}$ ${\rm M_{\odot}~{\rm yr^{-1}}}$
   (solid line), a factor of six below the unclumped H$\alpha$
   mass-loss rate (dotted line). For HD 93129A we show the
   $\tau_{\ast}$ values (right) along with the best-fit mass-loss rate
   model (solid line; $\dot{M} = 6.8 \times 10^{-6}$ ${\rm
     M_{\odot}~{\rm yr^{-1}}}$) and for comparison, the traditional
   unclumped H$\alpha$ mass-loss rate model.  The opacity model used
   for HD 93129A assumes altered CNO abundances (\cite{Taresch1997}).
 }
   \label{fig:9Sgr}
\end{center}
\end{figure}

There are nine lines and line complexes in the 9 Sgr {\it Chandra}
grating spectrum with high enough signal-to-noise for line profile
modeling to provide meaningful constraints. We assume a wind velocity
parameter $\beta = 0.7$ and find a mean shock onset radius of $R_{\rm
  o} = 1.4$ R$_{\ast}$, consistent with the EWS scenario. The ensemble
of $\tau_{\ast}$ values can be fit, given a model of the bulk wind
opacity (which we calculate assuming solar abundances), to derive a
mass-loss rate.  We find a mass-loss rate of $\dot{M} = 3.4 \times
10^{-7}$ ${\rm M_{\odot}~{\rm yr^{-1}}}$, which represents a factor of
six reduction over the traditional mass-loss rate derived from
H$\alpha$ and radio free-free emission, assuming a smooth wind
(\cite{ll1993, Puls1996}). In Fig.\ \ref{fig:9Sgr} we show the $R_{\rm o}$ and
$\tau_{\ast}$ results.

The O2 If* star, HD 93129A, is the earliest O star in the Galaxy and, according to
\cite[Taresch et al.\ (1997)]{Taresch1997}, has the highest mass-loss
rate of any O star, with $\dot{M} = 1.8 \times 10^{-5}$ ${\rm
  M_{\odot}~{\rm yr^{-1}}}$. More recent modeling (though also
ignoring clumping effects) gives $\dot{M} = 2.6 \times 10^{-5}$ ${\rm
  M_{\odot}~{\rm yr^{-1}}}$ and a wind terminal velocity of
$v_{\infty} = 3200$ km s$^{-1}$ (\cite{Repolust2004}). This extremely
strong and dense stellar wind provides an interesting test of the EWS
scenario for X-ray production in O stars.  Indeed, the {\it Chandra}
spectrum is quite hard, which if assumed to be due to high temperature
would make an EWS interpretation implausible.  However, the H-like Si
line strength is very weak, compared to the He-like Si line strength,
indicating a plasma emission temperature of no more than 8 million K,
which is consistent with LDI simulations of wind shocks. The hardness
of the X-ray spectrum appears instead to be due to severe attenuation
of the soft X-ray emission by both the interstellar medium and the
star's own wind.

Because of the absent soft X-rays, there are only four lines and line
complexes in the {\it Chandra} grating spectrum available for fitting.
We show the $\tau_{\ast}$ results in Fig.\ \ref{fig:9Sgr} with the
mass-loss rate fit superimposed.  For this star, too, we find a modest
mass-loss rate reduction of roughly a factor of four over the value
derived from H$\alpha$ fitting assuming no clumping.

\section{Broadband X-ray properties}

\begin{figure}[b]
\begin{center}
 \includegraphics[angle=90, scale=0.20]{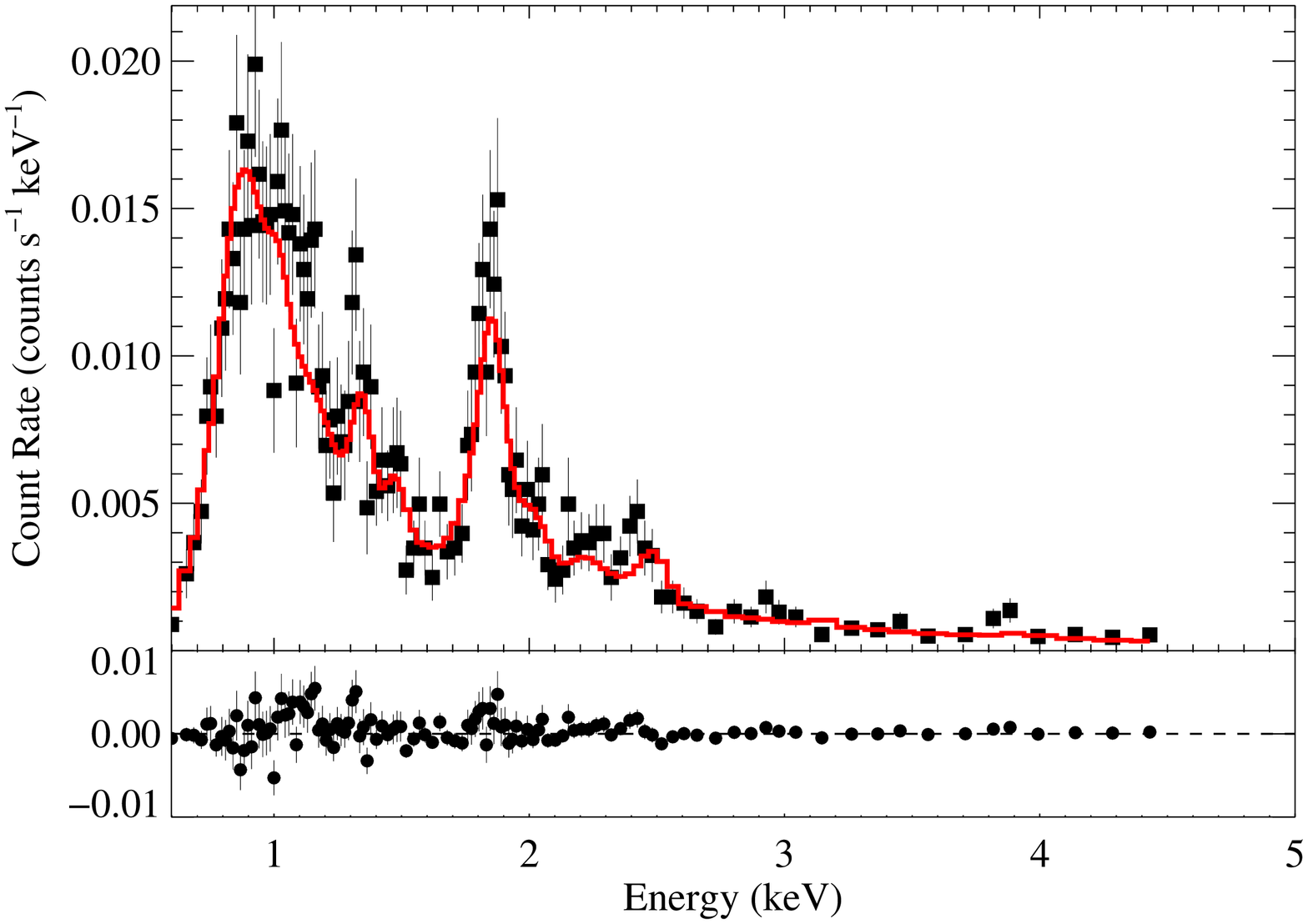} 
 \includegraphics[angle=90, scale=0.17]{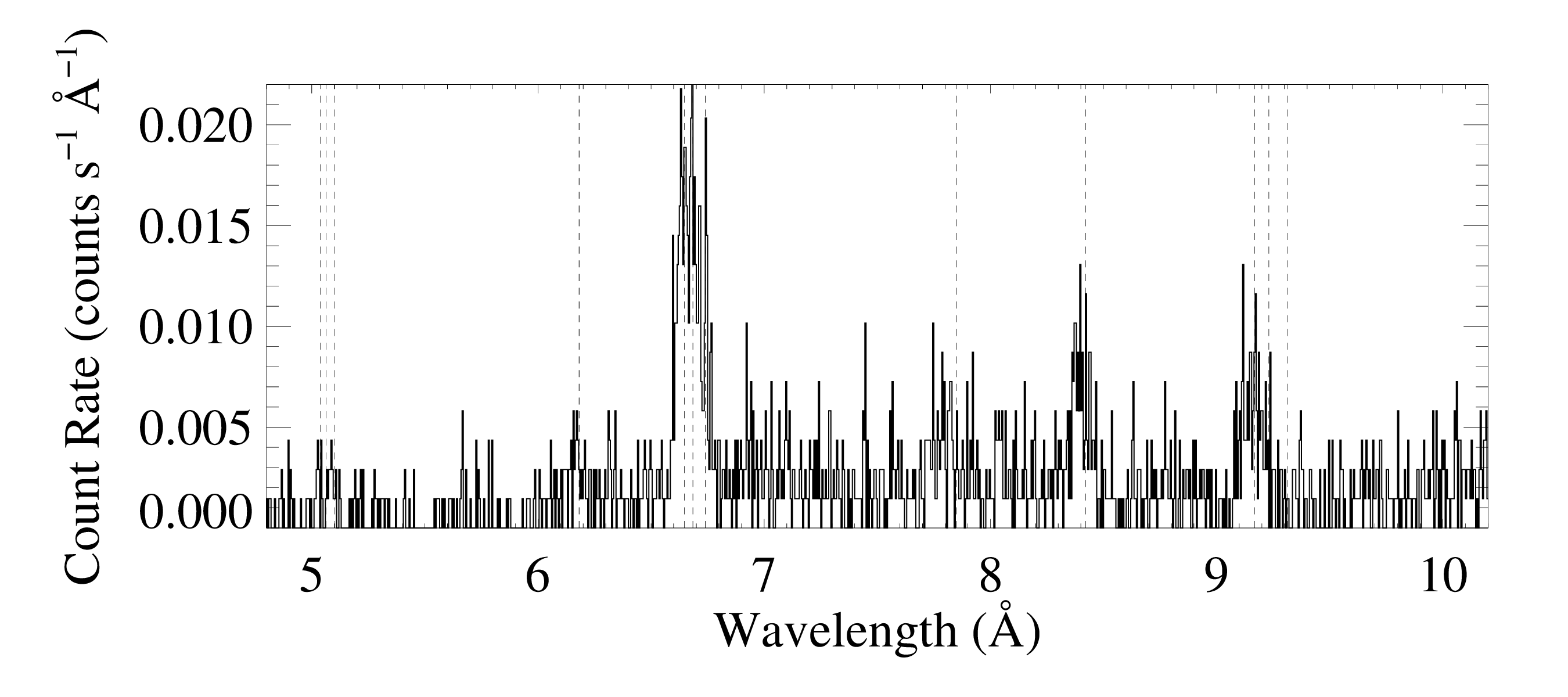} 
 \caption{The zeroth-order, low-resolution {\it Chandra} spectrum of
   HD 93129A, with the best-fit thermal emission model attenuated by
   stellar wind absorption (left).  This modeling shows that the
   emission temperature of the plasma is relatively low -- k$T = 0.6$
   keV -- while the effects of wind attenuation are significant;
   explaining the observed hardness.  The grating spectrum of this
   star is shown in the right-hand panel (dashed lines indicate rest
   wavelengths of important emission lines).  Note that while the
   spectrum is quite hard (no significant emission longward of 10
   \AA), the Si\, {\sc xiv} Ly$\alpha$ line at 6.18 \AA\/ is very weak
   compared to the Si\, {\sc xiii} complex near 6.7 \AA.  This very
   low H-like/He-like line ratio is indicative of plasma with a
   temperature of no more than 8 million K (0.7 keV). }
   \label{fig:hd93129_broadband}
\end{center}
\end{figure}

Given the strong wind absorption in the X-ray spectra of early O
stars, we have modeled the broadband spectral energy distributions
using simple one- and two-temperature thermal emission spectral models
(e.g.\ {\sc apec} \cite[Smith et al.\ (2001)]{Smith2001}) along with
wind attenuation, using the newly published radiation transport model,
{\it windtabs} (\cite{Leutenegger2010}).  This model accounts for the
spatial distribution of the emitting plasma within the absorbing wind,
and thus has a much more gradual decrease of transmission vs.\
fiducial optical depth than is seen in the exponential absorption
model that describes an absorbing medium in between the background
emitter and the observer, such as those employed in interstellar
absorption models.  The {\it windtabs} absorption model also uses a
realistic photoionization opacity model that includes partially
ionized metals and fully ionized H and He.  We fit the {\it Chandra}
zeroth-order spectrum (a CCD low-resolution spectrum) of HD 93129A
with this {\sc apec} and {\it windtabs} model and find a low plasma
temperature of 0.6 keV -- fully consistent with the LDI simulation
results -- and a significant wind column density, corresponding to a
mass-loss rate of $\dot{M} = 8 \times 10^{-6}$ ${\rm M_{\odot}~{\rm
    yr^{-1}}}$, which is consistent with the value we find from
fitting the individual line profiles (shown in the third panel of
Fig.\ \ref{fig:9Sgr}). We show the HD 93129A zeroth-order spectrum and
best-fit {\sc apec} and {\it windtabs} model in Fig.\
\ref{fig:hd93129_broadband}, along with the grating spectrum of the
star, in which the low H-like/He-like line ratios, indicative of low
plasma temperatures, can be seen.

\section{Conclusions}

The X-ray emission from normal massive stars can be understood in the
context of embedded wind shocks due to the line-driving instability.
Specifically, the X-ray emitting plasma shares the same kinematic
profile as the bulk wind. It is spatially distributed throughout the
wind above an onset radius of roughly $R_{\rm o} = 1.5$ R$_{\ast}$ and
-- from broadband modeling -- the plasma temperatures are less than 10
million K, in accord with the predictions of LDI simulations of EWSs.
These relatively low temperatures can be reconciled with the
relatively hard observed spectra by taking wind attenuation of the
soft X-rays into account.  When we model the effect of wind
attenuation on individual emission lines, we find that their modestly
blue-shifted and asymmetric profiles can be reproduced using mass-loss
rates that are lower by a factor of 3 to 6 compared to traditional
mass-loss rates that ignore clumping (and are consistent with newer
determinations that account for the clumping).  And furthermore, we
find that when we model the broadband spectral properties and account
for the effects of wind attenuation using a realistic radiation
transport model in conjunction with a realistic opacity model, we
derive similar mass-loss rate values.

Finally, we note that in this short paper we do not have the space to
discuss in detail the possible role of {\it porosity} in generating
the only modestly blue-shifted and asymmetric profiles.  Porosity
arises from clumping on very large scales, where individual clumps are
optically thick to X-ray photoelectric absorption (\cite{ofh2006}).
Our modeling suggests that porosity does not need to be invoked in
order to explain the observed X-ray properties of the early O stars we
discuss here.  Their properties are well explained by modest mass-loss
rate reductions.  Furthermore, porosity requires clumping, by
definition (but not the other way around).  So, once clumping is
invoked, and the density-squared diagnostics are adjusted accordingly,
there is no longer any need to invoke porosity to explain the data.
This and other aspects of porosity are addressed in the end-of-session
discussion, later in these proceedings.

\end{document}